# Full-scale field-free spin-orbit switching of the CoPt layer grown on vicinal substrates.

Luo, Yongming[1]; Liang, Mengfan[1]; Feng, Zhongshu[1]; Chen, Haoran[2]; Jiang, Nan[2]; Chen, Jianhui[1]; Yuan, Mingyue[2,3]; Zhang, Jingcang[3]; Cheng, Yifeng[3]; Sun, Lu[4]; Bai, Ru[1]; Miao, Xiaohe[5]; Wang, Ningning[1]; Wu, Yizheng[2]*; Che, Renchao[2,3]*

1. Center for Integrated Spintronic Devices, Hangzhou Dianzi University, Hangzhou 310018, China
2. State Key Laboratory of Surface Physics and Department of Physics, Fudan University, Shanghai 200433, China
3. Research Center for Intelligent Computing Platforms, Zhejiang laboratory, Hangzhou, 311121, China
4. School of Information Science and Technology, ShanghaiTech University, Shanghai 201210, China
5. Instrumentation and Service Center for Physical Sciences, Westlake University, Hangzhou 310024, China

## Abstract

A simple, reliable and field-free spin orbit torque (SOT)-induced magnetization switching is a key ingredient for the development of the electrical controllable spintronic devices. Recently, the SOT induced deterministic switching of the CoPt single layer has attracts a lot of interests, as it could simplifies the structure and add new flexibility in the design of SOT devices, compared with the Ferromagnet/Heavy metal bilayer counterparts. Unfortunately, under the field-free switching strategies used nowadays, the switching of the CoPt layer is often partial, which sets a major obstacle for the practical applications. In this study, by growing the CoPt on vicinal substrates, we could achieve the full-scale (100% switching ratio) field-free switching of the CoPt layer. We demonstrate that when grown on vicinal substrates, the magnetic easy axis of the CoPt could be tilted from the normal direction of the film plane; the strength of Dzyaloshinskii–Moriya interaction (DMI) would be also be tuned as well. Micromagnetic simulation further reveal that the field-free switching stems from tilted magnetic anisotropy induced by the vicinal substrate, while the enhancement of DMI help reducing the critical switching current. In addition, we also found that the vicinal substrates could also enhance the SOT efficiency. With such simple structure, full-scale switching, tunable DMI and SOT efficiency, our results provide a new knob for the design SOT-MRAM and future spintronic devices.

## Introduction:

Electrical manipulation of the magnetization is a key ingredient for future ultra-low-power, nonvolatile memory and logic devices. Spin−orbit torque (SOT) induced magnetization switching is one of the most competing techniques[1,2]. Through spin Hall effect or Rashba-Edelstein effects, SOT convert the charge current to spin current, which exert a torque on the magnetization.

Compared with the traditional spin transfer torque (STT) based devices, SOT show clear advantages in terms of devices duration [3], reduced power consumption [4] and faster device operation[5]. In this field, deterministic switching of the perpendicular magnetization has been one of the research focuses [14]. As the polarization direction of the spin current induced by spin Hall effect (SHE) or Rashba-Edelstein effects is in-plane, SOT is symmetric for the upward and downward magnetization, an auxiliary bias magnetic field is necessary for deterministically switching of a perpendicular magnetization by SOT [6], [7,8]. The requirement of the additional on-chip field component sets a major obstacle for achieving the full electrical manipulation of the magnetization. In recent years, different approaches have been proposed for the realization of field-free switching. For examples, by creating a build-in effective field such as exchange bias field, et. al [9,10], [11,12]; designing geometry asymmetry such as wedge [13] or other patterns [14]; creating composition gradient in the film structures [15],[16,17]; using unconventional out-of-plane polarized spin current [18-20].

Besides of the SOT induced switching in ferromagnet/Heavy metal (FM/HM) bilayer systems, the SOT induced magnetization switching CoPt single layer, which have attract a lot of recent interests [15,21], where a self-induced bulk-like SOT generated in the FM layer could induce the switching of the FM layer. Compared with the conventional interfacial-SOT in FM/HM bilayer systems, which maximizes the torque strength only for a very thin ferromagnet, the self-induced bulk SOT in the FM single layer could break this restriction which simplify the device structures and add new flexibility in device application. However, the switching performance is still need to be further improved in such single FM systems. However, by using conventional field-free switching strategies, such as interlayer exchange coupling [15], in-plane remanent magnetization [22], or crystal-symmetry design [21], the switching of the CoPt layer are often partial, the switching ratio is normally below 70%. As the switching directly determine the device performance in SOT-MRAM application [23,24], full-scale (100% switching ratio) and reliable switching of the CoPt single layer is highly demanded, while has not been reported to the best of our knowledge.

In this letter, by growing CoPt on vicinal $Al_2O_3$ substrates, we realize the field-free full-scale switching of the CoPt single layer. The switching has high durance, and have a wide operation window for CoPt grown on different vicinal substrates. We demonstrate that the vicinal substrates could induce the tilting of the magnetic anisotropy, and tune the strength of Dzyaloshinskii–Moriya interaction (DMI). Using micromagnetic simulation, we verify that the origin of the field-free switching is the tilted anisotropy, while DMI help reducing the critical switching current. In addition, we also found the SOT efficiency could be enhanced ~25% when grown on vicinal substrates. Due to the full scale, high endurance, simple structure and wide operation window, tunable DMI and SOT efficiency, our results provide an available design for simple, reliable and wafer scale application of SOT-MRAM and large-scale integrated circuit applications.

**Results**

## Structural and magnetic properties

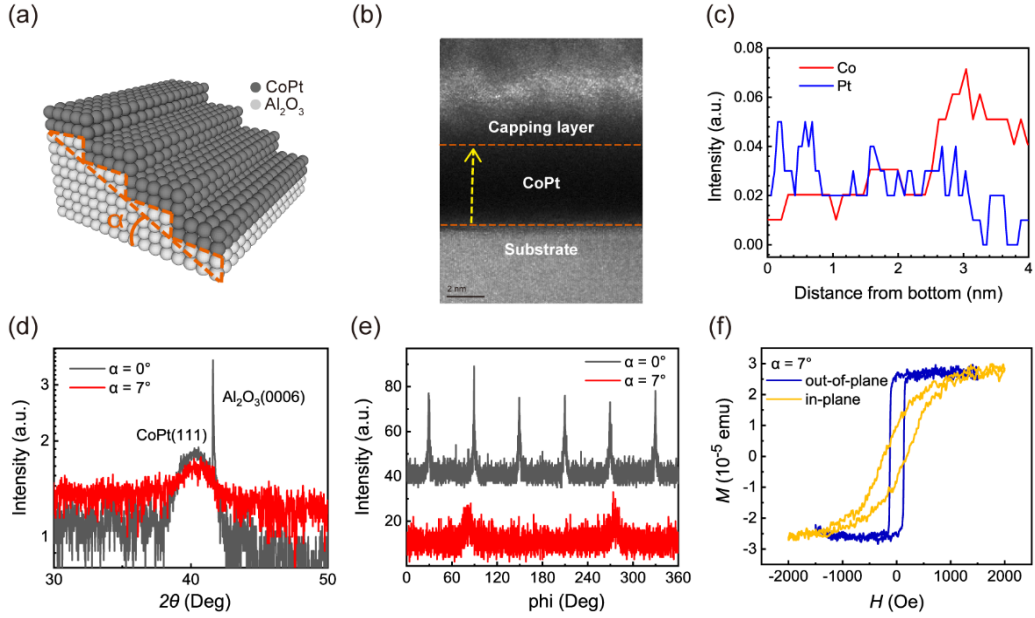

Figure1: (a) Schematic drawing of the CoPt film grown on vicinal substrates. (b) The high-resolution TEM image of the cross section of the CoPt film. (c) The line scanning of the EDS results marked by the yellow arrow in Fig.1(b). (d) and (e) are XRD spectra of the CoPt films grown on flat ($\alpha = 0°$) and vicinal substrate ($\alpha = 7°$). (d) The $2\theta$ line scan around the CoPt (111) and $Al_2O_3$ (0006)-diffraction condition. (e) Phi-scan pattern with CoPt (002) plane rotated along [111] axis. (f) In-plane and out-of-plane magnetic hysteresis loops for the CoPt layer deposited on vicinal substrate ($\alpha = 7°$).

Vicinal substrates denote those substrates with periodic atomic steps at the substrate surfaces. They are fabricated by inducing an inclination angle relative to the crystallographic plane during the cutting. In our experiment, we grow CoPt films on commercial-available vicinal $Al_2O_3$ (0001) substrates, which are schematically shown in Fig. 1(a). The CoPt films we grow exist an artificial composition gradient in the thickness direction, which are designed as the nominal multilayered structure of Pt(0.8)/Co(0.3)/Pt(0.6)/Co(0.5)/Pt(0.4)/Co(1) (thickness in nanometer). The vicinal angle of the substrates is denoted by $\alpha$ (shown in Fig. 1(a)). In our study, we use substrates with different $\alpha$ ($\alpha = 0°$ 5°, 7°, 10°). If not specified, the vicinal substrate represent here denotes substrate with $\alpha = 7°$, while $\alpha = 0°$ for the flat substrate. We use the transmission electron microscopy (TEM) to characterize the cross-section of film structure, as shown in Fig. 1(b), a smooth and continuous film of ~6 nm in thickness is obtained. The composition of Co and Pt in the film structure was verified by energy dispersive spectrometer (EDS), as shown in Fig. 1(c). The line scanning of the EDS results marked by the yellow arrow in Fig.1(b) revel that the Co and Pt elements exist an opposite composition gradient in the thickness direction, in accordance with our design of the film structure.

The crystal structure of the CoPt films could be verified by X-ray diffraction. Fig. 1(d) show the $2\theta$ line scan spectrums of the films deposited on flat ($\alpha = 0°$) and vicinal substrates

($\alpha = 7°$). For both samples we could observe a peak at 40.6°, this means that the CoPt films are (111) oriented, no matter the films are deposited on flat or vicinal substrates. The XRD peak also indicated that the multilayers have transformed to an alloying structure, due to the inter-diffusion. It should be noted that the $Al_2O_3$ (0006) substrate peak is only observed in the flat substrate, this may be due to the larger surface roughness on vicinal substrates, due to the existence of additional atomic steps. We further performed the off-axis Phi-scan for two different samples, as shown in Fig. 1(e). For the CoPt grown on flat substrate, the Phi-scan pattern with CoPt (002) plane rotated along [111] axis show six-fold symmetry with each diffraction peak separated by 60°. For CoPt with cubic symmetry, one should only observe three-fold rotational symmetry if viewed along the [111] direction [21]. The presence of the six diffraction peaks indicates that although it is (111) plane in z-direction, there is an additional type of twist domain in the material, where the domains are rotated along the substrate normal by 60°, similar with the situation when TiN (the same crystal structure type with CoPt) grown on $Al_2O_3$ (0001) substrates [25]. While for the CoPt grown on vicinal substrate, the Phi-scan pattern changes into two-fold symmetry. The reduction of the rotation symmetry indicates that the in-plane symmetry of the CoPt film is broken when grown on vicinal substrate. The two-fold Phi-scan pattern indicated that crystal structure of the CoPt film probably evolve from cubic to monoclinic, due to the anisotropic lattice strain that are parallel and perpendicular with the atomic steps on the vicinal substrates.

Fig. 1(f) shows the in-plane and out-of-plane magnetic hysteresis (M-H) loops for the CoPt layer deposited on vicinal substrate ($\alpha = 7°$). The results show that the in-plane and out-of-plane saturation fields are 1000 Oe and 190 Oe, respectively. The much larger in-plane saturation field and the square out-of-plane M-H loop indicate strong perpendicular magnetic anisotropy of the CoPt film. In supplementary information (SI) (Fig. S1), we further characterize the $M_s$ and the perpendicular anisotropy $H_k$ as a function of $\alpha$. The results reveal that for the films deposited on vicinal substrates, the additional strain from the substrates only slightly influence the value of $M_s$ and $H_k$.

**Full scale field-free switching**

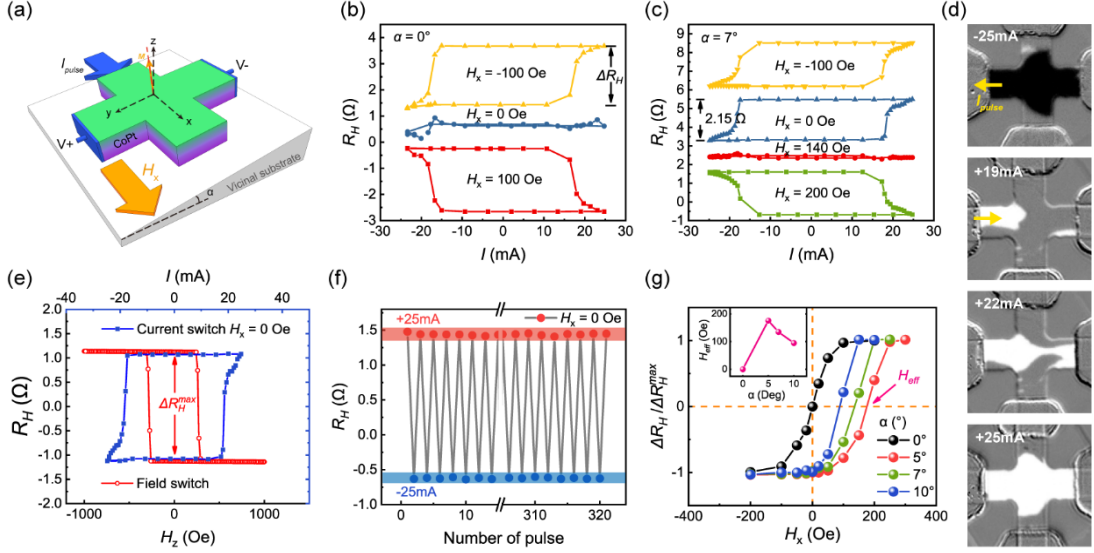

Figure 2: Current-induced switching of the CoPt layer. (a) Schematic drawing of the experiment set up. (b) SOT switching of the CoPt films grown on the flat substrate ($\alpha = 0°$), measured at $H_x = \pm 100$ Oe and 0 Oe. (c) SOT switching of the CoPt films grown on the vicinal substrate ($\alpha = 7°$). (d) Kerr images of the field-free switching process of the CoPt magnetization, for the films grown on vicinal substrates ($\alpha = 7°$). The bright domains and dark domains correspond to the up-ward and downward magnetization states, respectively. (e) Current and field ($H_z$) induced $R_H$ loops. Current-induced $R_H$ are measured with $H_x = 0$ Oe. (f) Field-free switching stability of the CoPt layer. Positive (+25mA) and negative (-25mA) current pulses are alternatively applied, the $R_H$ do not decay after ~300 switching cycles. (g) Switching ratio as a function of $H_x$, for CoPt films grown on substrates with different $\alpha$. Inset of (g) shows $H_{eff}$ (denoted by the pink arrows) as a function of $\alpha$.

Next, we characterize the SOT switching of the CoPt layer. The CoPt films grown on different substrates are patterned into Hall bar structures with 5 μm in width and 20 μm in length. Fig. 2 (a) is the schematical drawing of the experiment set-up. Pulse currents ($I_{pulse}$) with a fixed duration (30μs) are applied along X-direction. After each pulse, the magnetization states are electrically read out by the anomalous Hall resistance ($R_H = V_{ac}/I_{ac}$, $R_H \propto M_z$). The magnetization states could also be imaged by polar Kerr microscopy, which sensitive to the out-of-plane magnetization component. In-plane fields with different amplitudes $H_x$ are applied along the current direction, to help the deterministic switching of the magnetization. Fig. 2(b) shows the current induced switching of the CoPt layer that are grown on flat substrates. When the current pulses with different amplitudes are scanned, clear $R_H$ loops could be observed in the existence of $H_x$, but no loops when $H_x = 0$ Oe. The switching polarity are reversed with the direction of $H_x$. Such switching characteristics are in consistence with the typical SOT induced switching of the perpendicular magnetization [8,26,15], indicating the SOT origin of the CoPt switching. The results also demonstrated that the field-free switching could not be realized for the CoPt film deposited on flat substrate. Fig. 2(c) shows the switching behavior of the CoPt film grown on vicinal substrate ($\alpha = 7°$). As shown in Fig. 2(c), we also observe the reversal of switching polarity with the direction of $H_x$, indicating the same origin of the switching as that grown on flat substrates. The surprised results show that clear switching loops could be observed even in the absence of $H_x$, the critical $H_x$ for the reversal of the switching polarity shift to 140 Oe. The comparison of the switching performance demonstrating the field-free switching of the CoPt layer is induced by vicinal substrate. The field-free switching process could also be directly imaged by Kerr microscopy. As shown in Fig. 2(d), when the current reach +19 mA, the contrast of the Kerr image starts to change, indicating the switching of the magnetization, the critical switching current is in accordance with the current induced $R_H$ loop. Magnetic domains with reversed magnetization are first nucleated in the Hall bar, and then expand continuously through domain wall propagation with the increasing of the pulse current. The magnetization was fully switched at +25 mA. It should be noted that when grown on vicinal substrate, the Kerr images reveal that the field-free switching of the CoPt is full-scale (100 % switching ratio), which have not been realized previously. The switching ratio could be verified by current induced Hall resistance ($\Delta R_H$). As shown in Fig. 2(e), in the absence of the field, the current induced ($\Delta R_H$) is about $2.15\Omega$, which are nearly the same with the field induced maximum change of the $R_H$ ($\Delta R_H^{max} = 2.20\Omega$), demonstrating the switching of the CoPt magnetization is full-scale. We found that such field-free switching could only be realized when the current is perpendicular with the vicinal direction of the substrate (Y-axis in Fig. 2(a)), but could not when they are parallel, show in the SI (Fig. S2). We further test the switching endurance of our devices, we alternatively apply positive and negative current pulse with fixed amplitude (25 mA), which drive the CoPt switch back and forth continuously. As shown in Fig. 2(f), after ~300 cycles of switching, the Hall resistances at two different states exhibit negligible decay, demonstrating the high endurance of our system. Such high switching endurance also distinguish itself by using additional layers, such as antiferromagnetic layer or in-plane magnetic layers, where the magnetic structure are easily disrupted and result in the decay of the switching ratio [26].

We systematically characterize the switching behavior for devices grown on different vicinal substrates. We plot the switching ratio (defined as $\Delta R_H / \Delta R_H^{max} \times 100\%$) as a function of $H_x$, for films grown on different substrates (with different $\alpha$). The results are summarized in Fig. 2(g). Here the negative switching ratio denotes the opposite switching polarity. For the films deposited on flat substrate ($\alpha = 0°$), the curve is antisymmetric and we have $R_H(H_x) = -R_H(-H_x)$. The switching ratio increase with the external field and reach 100 % when $|H_x| > 100 Oe$. For all the samples grown on different vicinal substrates ($\alpha > 0°$), at zero field the switching ratio could reach nearly 100 %, demonstrating the wide operation window for such full-scale field-free switching. The critical field for zero switching shift from $H_x = 0$ Oe to nonzero values, which vary between 100~200 Oe, and is summarized in the inset of Fig. 2(g). These shift values reflect the effective in-plane field $(H_{eff})$ the vicinal substrates could produce, and is critical for the realization of the field-free switching. As summarized in Table. 1, the $H_{eff}$ are much larger than that produced by other strategies, which ensure the high switching ratio compared with other strategies.

## Quantitative Verification of the tilted anisotropy and DMI

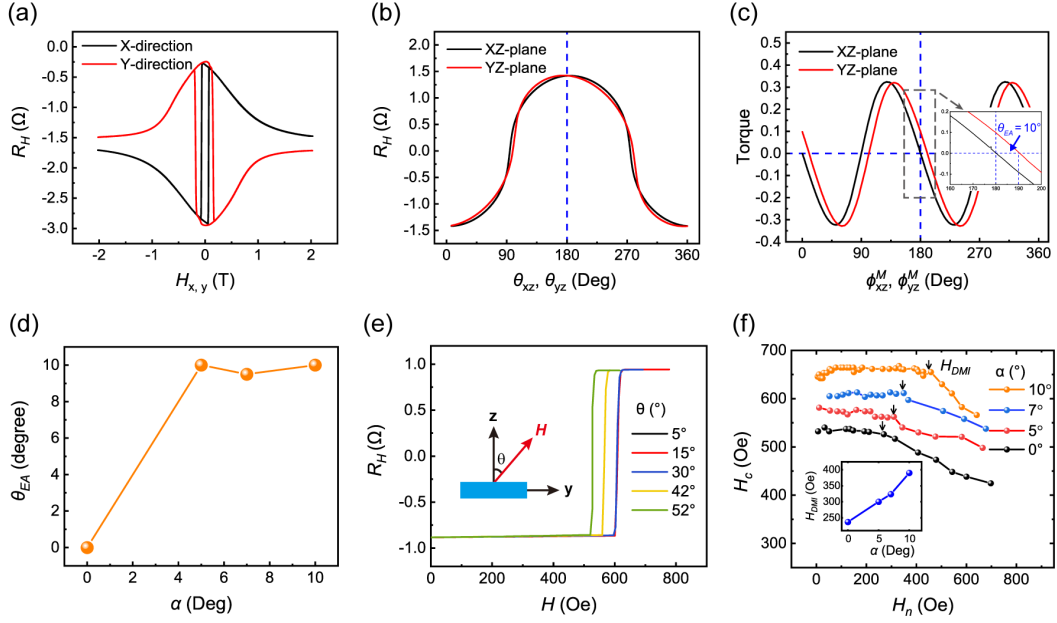

Fig. 3 (a) (b) and (c) are characterization of the tilted anisotropy for the device grown on vicinal substrates ($\alpha = 7°$), respectively. The Hall bar structure and the coordinates axis are the same as that is defined in Fig.2 (a). (a) Field-dependence of the $R_H$, with fields are scanned along X and Y directions. (b) Angular-dependence of the $R_H$, with the fields (1 T) are rotated in XZ and YZ planes. $\theta_{xz}, \theta_{yz}$ denotes the field angles with respect to the film normal direction (Z-axis) in the XZ and YZ planes. (c) Torque curves measured in the XZ and YZ planes. $\phi_{xz}^M, \phi_{yz}^M$ denote the magnetization angles with respect to the film normal direction (Z-axis) in the XZ and YZ planes. The inset is the enlarge of the data. (d) The tilting angle of EA as a function of $\alpha$. (e) $R_H$ curves when the magnetization switches from down to up under different magnetic field angles $\theta$ with respect to the film normal direction (as shown in the inset), for the sample grown on grown on vicinal substrates ($\alpha = 7°$). (f) $H_c$ as a function of $H_n$, for CoPt grown on different substrates. The inset shows the $H_{DMI}$ as a function of $\alpha$. $H_{DMI}$ denotes the threshold value of $H_n$, above which $H_c$ starts to decrease, denoted by the black arrow.

Next, we explore the physical origin of the field-free switching. According to previous results, both the tilted magnetic anisotropy[27] and Dzyaloshinskii–Moriya interaction (DMI) would play a critical role in the SOT the switching [17,28]. For this reason, we quantitatively characterize these quantities. When the CoPt grown on the vicinal substrates, the stray fields between adjacent atomic steps and the uniaxial strain could induce the magnetic easy-axis (EA) tilt from normal of the film plane. Such tilted EA could be verified by analyzing the rotation of the magnetization with respect to the film planes. In Fig. 3(a)-(c), for the SOT device grown on the vicinal substrate (the sample in Fig.2(c)), we characterize the field and angular dependence of the $R_H$, by applying a current (1 mA) in the X-direction and measuring the $R_H$ along the Y-direction. Fig. 3(a) show the field dependence of the $R_H$ when fields are scanned along X and Y directions. The field dependence of the $R_H$ curves along X and Y direction exhibits observable difference. The different $R_H$ curves along X and Y direction indicates that the EA is tilted from the normal direction of the film plane, or else the switching behaviors along different in-plane directions should be identical. To confirm the exact tilting plane, we further measure the angular dependence of the $R_H$ curves in the XZ and YZ plane. Fig. 3(b) shows the angular dependence of the $R_H$ curves when fields with fixed amplitude (1 T) are rotated in two different planes. The angular dependence of the $R_H$ along the XZ and YZ plane exhibits distinguishable difference. The $R_H$ curve is symmetric about Z-axis ($\theta_{xz} = 180°$) when rotated in the XZ plane, while asymmetric about Z-axis ($\theta_{yz} = 180°$) when rotated in the YZ plane. As the rotation of the magnetization should be symmetric bout the EA, the symmetric and asymmetric rotational symmetry about the film normal (Z-axis) demonstrate that the magnetization is tilted in the YZ plane (along the atomic steps of the substrates). Such tilted anisotropy induced field and angular dependence of the magnetization switching behavior could be reproduced by the micromagnetic simulation (shown in the SI, Fig. S4). For comparison, we also measure the sample that are grown on flat substrate (shown in the SI, Fig. S3). The results reveal that the field and angular dependence of $R_H$ along different directions are identical, in contrast with the sample grown on vicinal substrate. From the above discussion, we could conclude that when grown on vicinal substrates, the anisotropy of CoPt would tilt along the atomic steps of the substrates (Y- axis).

The exact tilting angle of the EA (defined as $\theta_{EA}$) could further be characterized by using the magnetic torquemetry [29]. In a system with uniaxial anisotropy ($K_u$), when a strong enough magnetic field $H$ is applied, energy density ($E/V$) could be expressed as:

$$E/V = -M_s H \cos(\theta - \phi^M) + K_u \sin^2(\phi^M) \qquad (1)$$

where $\theta$ is the angle between the applied magnetic field and the Z-axis, $\phi^M$ is the angle between the magnetization and the Z-axis, and could be obtained by the $R_H$ measurement ($\phi^M = \arccos(R_H / R_H^{\max})$). The equilibrium angle of the magnetization can be obtained by minimizing energy density with respect to $\phi^M$:

$$l(\phi^M) = H \sin(\theta - \phi^M) = (K_u / M_s) \sin 2(\phi^M) \qquad (2)$$

According to Eq. (2), the direction of EA corresponding to the direction where $l(\phi^M) = 0$. As shown in Fig. 3(c), for the CoPt grown on the vicinal substrate, we measure the $l(\phi^M)$ as a function of $\phi^M$ when the field (1 T) are rotated in the XZ and YZ planes. The results reveal there exists a phase difference for $l(\phi^M)$ measured in the XZ and YZ planes. When the field is rotated in the XZ plane, we have $l=0$ when magnetization is along Z-axis ($\phi_{xz}^M = 180°$), indicating that the EA is parallel the Z-axis in the XZ plane. However, in the YZ plane, the torque curve exist a 10° phase shift compared with that in the XZ plane, with $l=0$ when $\phi_{yz}^M = 190°$, this reflect that the EA tilt 10° from Z-axis in the YZ plane when grown on vicinal substrate. In the SI (Fig. S3), we also measure the $l(\phi^M)$ for the sample grown on flat substrate, the results shown that there do not exist such phase difference for $l(\phi^M)$ in XZ and YZ planes, and we have $l=0$ when $\phi^M = 180°$, demonstrating that EA is not tilted when grown on flat substrate. We also characterize the $\theta_{EA}$ for films grown on different substrates, as shown in Fig. 3(d). The results reveal that the tilting of EA is universal when grown on vicinal substrates.

Next, we further characterize the DMI of our CoPt layers, using a method proposed by Kim et al, which is based on the magnetic droplet nucleation model [30]. As depicted in the inset of Fig. 3(e), the hysteresis loop was measured by sweeping the magnetic field at an angle $\theta$ with respect to the z-axis. With the increasement of $\theta$, the switching field ($H_{sw}$) starts to decrease when $\theta$ reach certain critical value, as shown in Fig. 3(e). If we denote the magnetic field where the magnetization is switched from the "down" state to the "up" state by $H_{sw}$, then the coercive field $H_c$ and the accompanying in-plane field $H_n$ can be expressed by $H_{sw}\cos\theta$ and $H_{sw}\sin\theta$, respectively. Fig. 3(f) shows the measured $H_c$ as a function of $H_n$. As described in earlier works, the curve shows a clear plateau, and $H_c$ starts to decrease after $H_n$ passes a threshold value, which corresponds to $H_{DMI}$ [31]. By using this manner, we characterize the $H_{DMI}$ for films grown on different substrates. The results are shown in Fig. 3(f). For CoPt grown on flat substrate, $H_{DMI}$ is ~230 Oe, in consistence with the pervious results [15,32]. Our results also show a monotonously increasement of $H_{DMI}$ with $\alpha$, and reach ~390 Oe when $\alpha = 10°$. As previous results have shown that the strain from substrates would contribute a non-negligible DMI [33,34]. Our result further demonstrate that the additional strain induced by the periodic atomic steps on the vicinal substrates could tune and enhance the DMI strength.

**Simulation of the field-free switching**

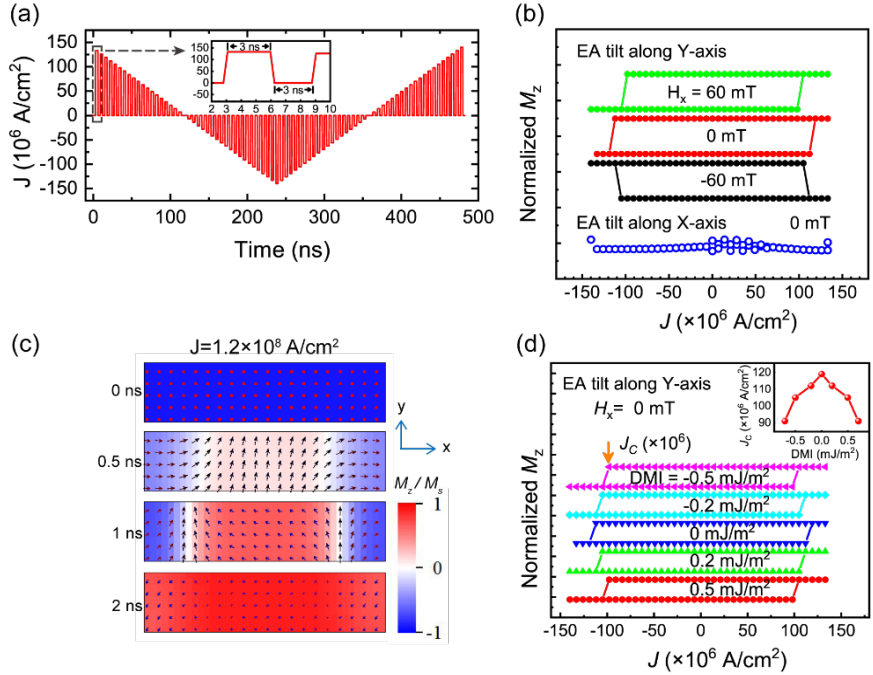

Figure 4: Simulation of the SOT switching of the CoPt. Currents are applied along X-direction. (a) Current pulse sequences used in the simulation. The inset shows the enlarge of the data, the pulse duration is 3 ns, and the pulse interval is 3ns. (b) SOT switching of CoPt films with tilted EA. The solid and hollow lines represent results with the EA tilt toward Y-axis and X-axis, respectively. The $\theta_{EA}$ are kept 10°. Fields $H_x$ are applied along the current directions. (c) The selected snaps of the magnetization during the switching. (d) SOT switching of CoPt layers with different DMI, in a system with tilted anisotropy. The inset shows the critical switching current densities as a function of DMI.

After characterizing the tilted anisotropy and DMI of our devices, we further verify the physical origin of the measured field-free switching, by using micromagnetic simulation. We consider the influence of the tilted anisotropy and DMI on the SOT induced magnetization switching. The simulation details are shown in the SI. We simulate the switching of a CoPt with $200\,nm \times 50\,nm \times 4\,nm\,(x \times y \times z)$ in size. We apply pulse currents with different amplitudes along X-direction, and record the magnetization states ($M_z$) after each pulse. The detail current sequences are shown in Fig. 4(a). External fields $H_x$ with different directions are applied along the current direction, to help the deterministic switching. Fig. 4(b) shows the switching of the CoPt layer with tilted anisotropy, with EA tilt toward different directions. When the EA is tilted along Y-axis (perpendicular with the current), clear switching loops could also be observed even at zero field, demonstrating the realization of the field-free switching in such switching geometry. While the EA is tilted along X-direction (parallel with the current), field-free switching could not be realized. Such switching behavior is in accordance with the experimental results in the SI (Fig. S2). Fig. 4(c) shows the detail switching process, a domain with reversed magnetization was first nucleated and then expanded towards the edge by domain wall motion, until the entire domain was reversed. Such domain nucleation and domain wall motion process are in accordance with the Kerr images in Fig. 2(d). Besides, we also simulated the effects of DMI on the field-free switching. Fig. 4(d) shows the influence of DMI on the SOT switching, in a CoPt layer with tilted anisotropy. The results reveal that the critical switching current decreasing and the increasement of DMI strength. This is because the DMI could inducing in-plane magnetization component at the edge of the hall bar structures. In CoPt system with perpendicular anisotropy, no matter with or without DMI (interface DMI, $0.5 mJ/m^2$), field-free switching also could not be realized, excluding the possibility of DMI induced field-free switching (shown in SI (Fig. S5)). The above discussion demonstrates that tilted anisotropy is the origin of the field-free switching, while DMI could help decreasing the critical switching current. Previous results also have shown that vicinal substrate could tilt the anisotropy of $L1_0$-FePt, and induce the field-free switching [27,35].

**Quantitative evaluation of the SOT effective field**

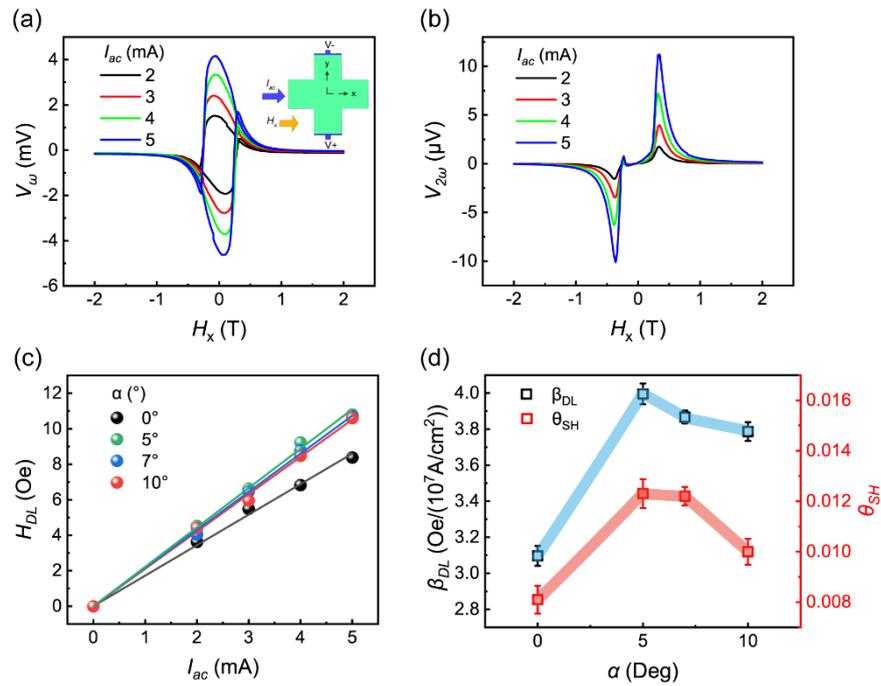

Fig. 5. Harmonic Hall voltage analysis of the CoPt films, for the films grown on different vicinal substrates. (a) and (b) are first and second harmonic Hall voltages as a function of $H_x$, using different currents $I_{ac}$ for the films deposited on vicinal substrates ($\alpha = 7°$). The inset shows the measurement geometry. (c) $H_{DL}$ as function of $I_{ac}$, for the CoPt films grown on different substrates. The dots and solid lines are experiment results and linear fitting of the data, respectively. (d) $\beta_{DL}$ and $\theta_{SH}$ as a function of $\alpha$.

We further quantitively characterize the SOT effective field and the SOT efficiency in our CoPt system, by using harmonic Hall voltage analysis. The measurement set up are schematically shown in the inset of Fig. 5(a), while an in-plane external magnetic field is swept (X-direction, with a small tilting angle 4° to the film plane), we measure the first ($V_\omega$) and second harmonic Hall voltages ($V_{2\omega}$) under an alternating (AC) current ($\omega = 13.7 Hz$) along X-direction. The second harmonic signal $V_{2\omega}$ is induced by the AC current, which exerts a periodic effective field on the magnetization. According to the harmonic analysis theory [36,37], the damping-like SOT effective field ($H_{DL}$), which is the main origin of the SOT switching, could be evaluated from fitting the field dependent $R_{xy}^\omega = V_{2\omega} / I_{ac}$ data using the following equation:

$$R_H^{2\omega} = \frac{R_H}{2} \frac{H_{DL}}{|H_x| - H_k^{eff}} + R_{offset} \qquad (3)$$

where $R_H$ is the anomalous Hall resistance, $H_{DL}$ and $R_{offset}$ are the damping-like SOT effective field and resistance offset, respectively. (The field-like SOT effective field, which is much weak than $H_{DL}$, could be evaluated when the fields are scanned along Y-direction (shown in Fig. S6 of SI). Fig. 5(a) (b) plot the field dependence of $V_\omega$ and $V_{2\omega}$ signals for the CoPt grown on the vicinal substrate ($\alpha = 7°$), measured under different $I_{ac}$ (2, 3, 4, and 5mA), from which $H_{DL}$ could be calculated according to Eq. 3. We did not consider the influence of planner hall effects, for its influence on the $R_{xy}^{2\omega}$ is negligible [15]. As shown in Fig. 5(c), $H_{DL}$ exhibit a linear dependence with the amplitude of AC current, demonstrating the current origin of $H_{DL}$. we further characterize the $H_{DL}$ for the CoPt deposited on different vicinal substrates ($\alpha$ = 0° 5°, 7°, 10°). For all the samples, $H_{DL}$ exist distinguishable difference for the samples grown on different substrates, indicating the influence of the substrates on the $H_{DL}$. To quantify such difference, we further calculate the SOT efficiency ($\beta_{DL}$) and spin Hall angle ($\theta_{SH}$) for the different samples. The SOT efficiency is defined as $\beta_{DL} = H_{DL} / j_{CoPt}$, which could be obtained by fitting the current dependence of the $H_{DL}$ in Fig. 5(c). The spin Hall angle could also be calculated as:

$$\theta_{SH} = \left(\frac{2e}{\hbar}\right) \mu_0 M_s t_{FM} \beta_{DL}$$

where e, $\hbar$ and $\mu_0$ are the elementary charge, reduced Planck constant and permeability of vacuum, respectively. $M_S$ and $t_{FM}$ are the saturation magnetization and thickness of the CoPt films. Fig. 5(d) show the $\beta_{DL}$ and $\theta_{SH}$ for films deposited on different vicinal substrates. For the films deposited on the flat substrate ($\alpha = 0°$), the calculated $\beta_{DL}$ = 3.1 Oe per $10^7$ A/cm², $\theta_{SH}$ =0.008. For the devices fabricated on vicinal substrates ($\alpha$ = 5°, 7°, 10°), our results show that the $\beta_{DL}$ and $\theta_{SH}$ have an enhanced of ~25%, compared with that grown on flat substrates. Thus, our results demonstrate that the vicinal substrate could not only induce field-free switching, but also increase the SOT efficiency. The tuning of the bulk SOT by vicinal substrate have also been observed in the

L1$_0$-FePt system [27]. This could be contributed to the strain effects, the strain induced by the additional atomic steps would change crystal symmetry and the modify the spin–orbit coupling, and orbital polarization [38].

**Discussion:**

| Methods | Structure | Switching ratio | H$_{eff}$ | Reference | Year |
|---|---|---|---|---|---|
| Interlayer exchange coupling (T-type) | IrMn/Co/Ru/CoPt | 50 % | 60 Oe | [15] | 2021 |
| Wedge structure | CoPt single layer | 67 % | 80 Oe | [32] | 2022 |
| Addition in-plane anisotropy | CoPt single layer | 40 % | 15 Oe | [22] | 2022 |
| Crystal asymmetry (3m torqe) | CoPt single layer | 60 % | Not provided | [21] | 2022 |
| **Vicinal substrate** | **CoPt single layer** | **100 %** | **190 Oe** | **This work** | |

Table. 1 Summary of the bulk-SOT induced field-free switching of the CoPt layer

Finally, we compare the field-free switching performance of the CoPt layer by using different strategies, as summarized in Table. 1. Previously, no matter by using interlayer exchange coupling, wedge-shape structures, additional in-plane anisotropy or crystal asymmetry, the field-free switching ratio is below 70 %. While in this work, the 100 % switching ratio could be realized. The high switching ratio could be attribute to the large effective field induced by vicinal substrates, which could reach up to 190 Oe in this work, while below 80 Oe by using other strategies. Besides of the switching ratio, our approach could also distinguish itself from the switching stability and simple design. Conventionally, to achieve the field-free switching, additional magnetic or antimagnetic layers are often necessary, and the switching are easily disrupted and result in the decay of the switching ratio, due the change of the magnetization states of the additional functional layers [26]. In addition, our method is also more suitable for wafer-scale production compared with the wedge structures, as the wafer-scale vicinal substrate is easy to obtain and is commercially available.

In summary, by growing CoPt on vicinal Al$_2$O$_3$ substrates, we realize the field-free full-scale switching of the CoPt single layer. We demonstrate that the vicinal substrates could induce the tilting of the magnetic anisotropy, and tune the strength of Dzyaloshinskii–Moriya interaction (DMI). We verify that the origin of the field-free switching is the tilted anisotropy, while DMI help reducing the critical switching current. The SOT efficiency of the CoPt could also be enhanced when grown on vicinal substrates. Due to the full scale, high endurance, simple structure and wide operation window, tunable DMI and SOT efficiency, our results provide an available route for designing simple and reliable SOT-MRAM and large-scale integrated circuit applications.

## Methods

**Sample growth and device fabrication.** The CoPt films were deposited on vicinal $Al_2O_3$ (0001) substrates by magnetron sputtering under a base pressure 3 x $10^{-8}$ Torr. To prevent the oxidation, a layer of $TaO_x$ (1.5) are capped on top of the film structure. The temperature was kept at room temperature during the deposition process. Vibrating sample magnetometry was used to quantify the magnetic properties of each sample.

**TEM sample preparation and characterization.** TEM samples of the CoPt films were fabricated by a focused ion beam machine. Microstructures were studied by transmission electron microscopy (TEM). Energy-dispersive X-ray spectroscopy (EDS) analysis was performed by using scanning transmission electron microscopy (STEM) mode.

**Device fabrication and electrical measurement.** The films were fabricated into Hall bar devices of 5 μm in width and 20 μm in length, by using photolithography and argon ion milling. For field and angular dependence of the $R_H$ ($R_H = V_{ac} / I_{ac}$) measurements, we apply an AC current (1 mA, 13.7 Hz, provided by Keithley 6221) in the X-direction, while measure the $R_H$ along the Y-direction, by using the SR830 lock-in amplifier. For SOT switching measurements, a pulse currents with a fixed duration of $30 \mu s$ are applied along X-direction. After each pulse, the magnetization states are electrically read out by the anomalous Hall resistance rough SR830 lock-in amplifier by applying a small AC current ($I_{ac}$ = 50 μA). In-plane fields with different amplitudes $H_x$ are applied along the current direction, to help the deterministic switching of the magnetization. For harmonics measurements, two SR830 lock-in amplifiers were used to detect the first and second harmonic Hall voltage induced by an AC current with a frequency of 13.7 Hz provided by Keithley 6221.

**Simulation.** Micro-magnetization simulation was conducted by using the OOMMF software. We simulate a CoPt system which is 200 nm x 50 nm x 4 nm in size, the unit cell size is 5 nm x 5 nm x 4 nm. The magnetic parameters of CoPt films as follows: the exchange stiffness: $A = 1 \times 10^{-11} J/m$; saturation magnetization: $M_s = 2 \times 10^5 A/m$; the perpendicular anisotropy is $K_u = 8 \times 10^4 J/m^3$, damping $\alpha = 0.01$.

## Acknowledgement:

We thank Dr. Tainping Ma, Dr. Juexue Li, Dr. Quanlin Ye, Dr. Qian Li and Dr. Gong Cheng for helpful discussion. This work is supported by the National Natural Science Foundation of China (Grants No. 12274108), Natural Science Foundation of Zhejiang Province (Grants No. LY23A040008 and No. LY23A040008) and Basic scientific research project of Wenzhou（Grants No. G20220025）.

# References

1. Shao, Q. et al. Roadmap of Spin‐Orbit Torques. *Ieee Trans. Magn.* **57**, 1-39 (2021).
2. Manchon, A. et al. Current-induced spin-orbit torques in ferromagnetic and antiferromagnetic systems. *Rev. Mod. Phys.* **91**, 35004 (2019).
3. Ramaswamy, R., Lee, J. M., Cai, K. & Yang, H. Recent advances in spin-orbit torques: Moving towards device applications. *Appl. Phys. Rev.* **5**, 31107 (2018).
4. Aradhya, S. V., Rowlands, G. E., Oh, J., Ralph, D. C. & Buhrman, R. A. Nanosecond-Timescale Low Energy Switching of In-Plane Magnetic Tunnel Junctions through Dynamic Oersted-Field-Assisted Spin Hall Effect. *Nano Lett.* **16**, 5987-5992 (2016).
5. Garello, K. et al. Ultrafast magnetization switching by spin-orbit torques. *Appl. Phys. Lett.* **105**, 212402 (2014).
6. Fukami, S., Anekawa, T., Zhang, C. & Ohno, H. A spin‐orbit torque switching scheme with collinear magnetic easy axis and current configuration. *Nat. Nanotechnol.* **11**, 621-625 (2016).
7. Miron, I. M. et al. Perpendicular switching of a single ferromagnetic layer induced by in-plane current injection. *Nature.* **476**, 189-193 (2011).
8. Liu, L. et al. Spin-Torque Switching with the Giant Spin Hall Effect of Tantalum. *Science.* **336**, 555-558 (2012).
9. Fukami, S., Zhang, C., DuttaGupta, S., Kurenkov, A. & Ohno, H. Magnetization switching by spin‐orbit torque in an antiferromagnet‐ferromagnet bilayer system. *Nat. Mater.* **15**, 535-541 (2016).
10. Kong, W. J. et al. Spin‐orbit torque switching in a T-type magnetic configuration with current orthogonal to easy axes. *Nat. Commun.* **10**, 233 (2019).
11. Lau, Y., Betto, D., Rode, K., Coey, J. M. D. & Stamenov, P. Spin‐orbit torque switching without an external field using interlayer exchange coupling. *Nat. Nanotechnol.* **11**, 758-762 (2016).
12. Zhao, Z., Smith, A. K., Jamali, M. & Wang, J. External-Field-Free Spin Hall Switching of Perpendicular Magnetic Nanopillar with a Dipole-Coupled Composite Structure. *Adv. Electron. Mater.* **6**, 1901368 (2020).
13. Yu, G. et al. Switching of perpendicular magnetization by spin‐orbit torques in the absence of external magnetic fields. *Nat. Nanotechnol.* **9**, 548-554 (2014).
14. Safeer, C. K. et al. Spin‐orbit torque magnetization switching controlled by geometry. *Nat. Nanotechnol.* **11**, 143-146 (2016).
15. Xie, X. et al. Controllable field-free switching of perpendicular magnetization through bulk spin-orbit torque in symmetry-broken ferromagnetic films. *Nat. Commun.* **12**, 2473 (2021).
16. Shu, X. et al. Field-Free Switching of Perpendicular Magnetization Induced by Longitudinal Spin-Orbit-Torque Gradient. *Phys. Rev. Appl.* **17**, 24031 (2022).
17. Wu, H. et al. Chiral Symmetry Breaking for Deterministic Switching of Perpendicular Magnetization by Spin‐Orbit Torque. *Nano Lett.* **21**, 515-521 (2021).
18. Hu, S. et al. Efficient perpendicular magnetization switching by a magnetic spin Hall effect in a noncollinear antiferromagnet. *Nat. Commun.* **13**, 4447 (2022).
19. Baek, S. C. et al. Spin currents and spin‐orbit torques in ferromagnetic trilayers. *Nat. Mater.* **17**, 509-513 (2018).
20. MacNeill, D. et al. Control of spin‐orbit torques through crystal symmetry in WTe2/ferromagnet bilayers. *Nat. Phys.* **13**, 300-305 (2017).

21. Liu, L. et al. Current-induced self-switching of perpendicular magnetization in CoPt single layer. *Nat. Commun.* **13**, 3539 (2022).
22. Li, Z. et al. Field-Free Magnetization Switching Induced by Bulk Spin‑Orbit Torque in a (111)-Oriented CoPt Single Layer with In-Plane Remanent Magnetization. *Acs Appl. Electron. Mater.* **4**, 4033-4041 (2022).
23. Wu, H. et al. Field-free approaches for deterministic spin‑orbit torque switching of the perpendicular magnet. *Materials Futures*. **1**, 22201 (2022).
24. Gallagher, W. J. Emerging nonvolatile magnetic memory technologies. 2010 10th IEEE International Conference on Solid-State and Integrated Circuit Technology. 2010 2010-1-1: IEEE; 2010. p. 1073-1076.
25. Yu, W. et al. Epitaxial titanium nitride microwave resonators: Structural, chemical, electrical, and microwave properties. *Phys. Rev. Mater.* **6**, 36202 (2022).
26. Liu, L., Lee, O. J., Gudmundsen, T. J., Ralph, D. C. & Buhrman, R. A. Current-induced switching of perpendicularly magnetized magnetic layers using spin torque from the spin Hall effect. *Phys. Rev. Lett.* **109**, 96602 (2012).
27. Luo, Y. et al. Field-free switching through bulk spin—orbit torque in L10-FePt films deposited on vicinal substrates. *Front. Phys.* **17**, 53511 (2022).
28. Zheng, Z. et al. Field-free spin-orbit torque-induced switching of perpendicular magnetization in a ferrimagnetic layer with a vertical composition gradient. *Nat. Commun.* **12**, 4555 (2021).
29. Xiao, X. et al. Magnetic anisotropy in Fe films epitaxied by thermal deposition and pulse laser deposition on GaAs(001). *J. Appl. Phys.* **113**, 17C114 (2013).
30. Kim, S. et al. Magnetic droplet nucleation with a homochiral Néel domain wall. *Physical Review. B*. **95**, 220402 (2017).
31. Kim, D. et al. Bulk Dzyaloshinskii‑Moriya interaction in amorphous ferrimagnetic alloys. *Nat. Mater.* **18**, 685-690 (2019).
32. Huang, Q. et al. Field-Free Magnetization Switching in a Ferromagnetic Single Layer through Multiple Inversion Asymmetry Engineering. *Acs Nano*. **16**, 12462-12470 (2022).
33. Xia, S. Y. et al. Source and origin of the interfacial Dzyaloshinskii-Moriya interaction in a heavy-metal|magnetic-insulator bilayer. *Phys. Rev. B*. **105**, 184417 (2022).
34. Xu, Z. et al. Strain-Tunable Interfacial Dzyaloshinskii‑Moriya Interaction and Spin-Hall Topological Hall Effect in Pt/$Tm_3Fe_5O_{12}$ Heterostructures. *Acs Appl. Mater. Interfaces*. **14**, 16791-16799 (2022).
35. Tao, Y. et al. Field-free spin‑orbit torque switching in L10-FePt single layer with tilted anisotropy. *Appl. Phys. Lett.* **120**, 102405 (2022).
36. Hayashi, M., Kim, J., Yamanouchi, M. & Ohno, H. Quantitative characterization of the spin-orbit torque using harmonic Hall voltage measurements. *Phys. Rev. B*. **89**, 144425 (2014).
37. Qiu, X. et al. Angular and temperature dependence of current induced spin-orbit effective fields in Ta/CoFeB/MgO nanowires. *Sci. Rep.* **4**, 449 (2014).
38. Filianina, M. et al. Electric-Field Control of Spin-Orbit Torques in Perpendicularly MagnetizedW/CoFeB/MgO Films. *Phys. Rev. Lett.* **124**, 217701 (2020).